\documentclass[aps]{revtex4}
\usepackage{amsmath}
\usepackage{amssymb}
\usepackage{graphicx}

\DeclareMathOperator{\arccot}{arccot}

\begin{document}

\title{Symmetry-based estimation of lower bound on secure key rate of noisy private states}

\author{Jan Tuziemski}

\affiliation{Faculty of Applied Physics and Mathematics,
Technical University of Gda\'nsk, PL-80-952 Gda\'nsk, Poland}

\affiliation{National Quantum Information Centre of Gda\'nsk,
PL-81-824 Sopot, Poland}

\author{Pawe\l{} Horodecki}

\affiliation{Faculty of Applied Physics and Mathematics,
Technical University of Gda\'nsk, PL-80-952 Gda\'nsk, Poland}

\affiliation{National Quantum Information Centre of Gda\'nsk,
PL-81-824 Sopot, Poland}

\begin{abstract}
Quantum private states are the states that represent some amount of perfect secure key. 
A simple symmetry of any generalised private quantum state (ie. the states that represent
perfect key but not fully random) is provided and extended on 
Devetak-Winter so called ccq (classical-classical-quantum) 
and cqq (classical-quantum-quantum) lower bound on secure key.
This symmetry is used to develop a practical method  of estimating the 
Alice measurement that is optimal form the perspective of single shot 
Devetak-Witner lower bound on secure key. The method is particularly good when 
the noise does not break the symmetry of the state with respect to 
the lower bound formula. It suggest a general paradigm 
for quick estimation of quantum communication rates under the symmetry 
of a given resource like state and/or channel.
\end{abstract}



\date{\today}

\maketitle

\pagenumbering{arabic}

\section{Introduction}
Entanglement is considered as a resource in quantum communication 
and computing. It has many intriguing properties that make in some cases quantum physics predictions
drastically different from classical ones (see \cite{reviews}).
Any BB84 type protocol 
\cite{BB84} is formally equivalent to some version 
of entangled based protocol of the type E91 \cite{Ekert}. On the level of uncorrelated sources
Bennett et al. pointed out this fact on specific scheme \cite{BBM} which was later naturally extended 
to a quantum privacy amplification QPA \cite{QPA} based
entanglement distillation protocol \cite{Bennett-distillation}. The general intrinsic connection between BB84 secret key generation
and  possibility of maximal entanglement distillation from correlated
sources of noisy entanglement was provided explicitly by elegant
error correction type analysis \cite{SP2000} which in particular
illuminated this aspect hidden in previous proofs.
However the fundamental intuition behind the  BB84
secret key generation - entanglement distillation equivalence
is already present in the case of uncorrelated source
for which QPA protocol works. The latter is a protocol that distills  maximally entangled 
states out of a mixed states in a well defined way.
In QPA it is eavesdropper that is representing the noise 
and the distillation procedure is aimed at remove the correlations 
with eavesdropper in the process that produces a pure output state 
- maximally entangled state that is a source of perfect key. 
It seemed that this QPA is necessary to get privacy.  However there exists 
nondistillable entanglement called bound entanglement \cite{bound} for which 
by the very definition QPA  in its original form can not work.
It turns out that there is another possibility of distilling secret key 
by distilling private states \cite{secret-key-bound} that are generalisations of maximally 
entangled states - they provide secret key under the measurement 
in a fixed basis on some part of the system. 
The measurement basis may be unique and that is what makes the private 
states more general form maximally entangled ones for which there 
are infinite many pairs of local bases that provide secure key.
The distillation of private states allow to provide secret key form 
bound entanglement (see \cite{secret-key-bound}) showing in particular 
a possibility of drastic separation between amount of 
pure entanglement that can be distilled from a quantum state 
(called distillable entanglement and denoted by D) 
and  amount of secret key that can be distilled from 
the same state (called disiilalble secret key and denoted by K).
Recently the separation $D < K$ was experimentally  
demonstrated together with the illustration how inefficient 
may be the original entanglement distillation based scheme, 
if compared with  p-bit based protocols \cite{Dobek-Et-Al}.

In this article we present the new symmetry of the states with perfect secure key called generalized private states, which states that in the most popular Devetak-Winter protocol secret key rate is invariant in both scenarios of CCQ and CQQ type if the measurement bases, chosen in a wrong way, are rotated around the axis corresponding to the secure basis by any angle. Then we use this symmetry to propose a new method of searching the optimal basis, which allows to obtain optimal amount of distillable key with respect to the Devetak-Winter secret key rate, in the case when a state possessing this symmetry was rotated by an unknown angle. We investigate the influence of various qubit channels on the result of the method and show that the proposed method is an optimal one as long as a channel is bistochastic. Finally we derive the error estimation of numerical implementation of the procedure and show examples of the results.

\section{A useful symmetry of the generalised private states}

According to \cite{hhho2009} any state containing 
perfectly secure key 
corresponds to a pure state shared by three parties  Alice, Bob and Eve. 
Unlike the eavesdropper subsystem $E$ with  a Hilbert space 
${\cal H}_{E}$ the subsystems of Alice and Bob 
are composite and correspond to the tensor products of 
Hilbert spaces ${\cal H}_{A} \otimes {\cal H}_{A'}$
and  ${\cal H}_{B} \otimes {\cal H}_{B'}$ respectively.
We call the subsystem corresponding to the pair A,B the key part since this is the 
part that is used for key generation by local von Neumann measurements 
while the pair A'B' is called the shield part since it is in 
a sense responsible for protecting the key.
The above structure allows to write explicitly the pure state of the three parties 
which represents a perfectly secure random 
a perfectly secure random statistics  $\vec{p}=[p_0,...,p_{d-1}]$. 
It is a pure state $|\Psi_{AA'BB'}\rangle $ of the following form which we shall 
call {\it generalised private states}:
\begin{equation}
|\Psi_{AA'BB'E}\rangle = \sum_{ij} c_{i}c_{j} |ii\rangle_{AB} \otimes [U_{A'B'}^{(i)} \otimes I_E]|\Psi_{A'B'E} \rangle
\label{p-bit-pure}
\end{equation} 
for some fixed $|\Psi_{A'B'E} \rangle$, some unitaries $U_{A'B'}^{(i)}$ and 
probabilities $\{p_{i}=|c_{i}|^{2}\}$. The basis $|ii \rangle_{AB}$ is called 
the secure basis since after performing the local von Neumann measurements 
in that basis Alice and Bob share the correlated probability distribution 
$\{ p_{AB}^{ij}=\delta_{ij}p_i \}_{i=0}^{d-1}$ which is completely uncorrelated form the 
system E. Here we do not assume them to be necessarily $p_i=\frac{1}{d}$ as it is in the 
case of the {\it private states}. In fact the density matrix corresponding 
to the state vector (\ref{p-bit-pure}) is of the form: 
\begin{equation}
\rho_{priv} =\sum_{ij} c_{i}c_{j}[\left|ii\right\rangle \left\langle jj\right|]_{AB} \otimes [\left|\Psi_{i}\right\rangle \left\langle \Psi_{j}\right|]_{A'B'E},
\label{p-bit-mixed}
\end{equation}
where we put $|\Psi_i\rangle=[U_{A'B'} \otimes I_{E}]|\Psi\rangle_{A'B'E}$ dropping the 
superscripts $A'B'E$.
If Alice and Bob measure the subsystems $AB$ in some basis $\{|e_i\rangle_{A}\}$, $\{|f_i\rangle_{B}\}$
and trace the shield subsystems $A'B'$ they get the so called 
{\it the form of the CCQ state with respect to the (product) basis ${\cal B}_{AB}=\{|e_i f_i\rangle_{AB}
\equiv |e_i \rangle_{A} \otimes |f_i\rangle_{B}\}$}:
\begin{equation}
\rho^{CCQ,{\cal B}_{AB}}_{ABE}(\{|e_i, f_j\rangle \})=\sum^{d-1}_{i=0} q_i \left|e_i\right\rangle \left\langle e_i\right|_A 
\otimes \left|f_i\right\rangle \left\langle f_i\right|_B \otimes\rho^i_{E} \nonumber
\end{equation}
with some probability distribution $q_i$.
It happens that if they choose the basis $\{|e_i f_j \rangle_{AB} \}$ to be just equal 
to the secure one ${\cal B}_{AB}^{0}=\{|ij\rangle_{AB}\}$ then the above state reduces to the product form 
\begin{equation}
\rho^{CCQ,{\cal B}^{0}_{AB}}_{ABE}=(\sum^{d-1}_{i=0} p_i [\left|ii\right\rangle \left\langle ii\right|]_{AB} ) \otimes \rho_{E} \nonumber
\label{product} 
\end{equation}
with $q_i=p_i$. Because of the explicitly product form - no correlations of the E system 
with  the key part AB are present here. 
Before proving some new property let us remind that the so called CQQ state with respect to the basis ${\cal B}_{A}=\{|e_i\rangle_{A}$
which results form Alice local von Neumann measurement and tracing out both A'B':
\begin{eqnarray}
\rho^{CQQ,{\cal B}_{A}}_{ABE}=\sum^{d-1}_{i=0} q_i \left|e_i\right\rangle \left\langle e_i\right|_A \otimes \rho^i_{BE} 
\end{eqnarray}

Note that measuring the private state in the local basis ${B}_{A}^{0}$ being just the reduction of 
the product secure basis ${\cal B}^{0}_{AB}$ we get still the CCQ state in the form (\ref{product}) 
rather than the general CQQ state (note that any CCQ state is a CQQ one but not vice versa) which is a consequence 
of the private character of the state. 

\subsection{The Devetak-Winter protocol rates}

We have a natural definition of the key rates in one-way protocols 
obtained by measuring the state first in some local basis ${\cal B}_{A}$ 
or a product one ${\cal B}_{AB}$ which will produce the 
CQQ or CCQ state respectively and then calculating the difference of the 
Holevo functions of the states:
\begin{equation}
K_{DW}^{{\cal B}_{A}}(\rho_{ABE})=I_{A:B}(\rho_{AB}^{CQ}) - I_{A:E}(\rho_{AE}^{CQ})
\label{K-CQQ}
\end{equation}
with $\rho_{AB}^{CQ}$, $\rho_{AE}^{CQ}$ being a suitable reductions of the 
state $\rho_{ABE}^{CQQ,{\cal B}_{A}}$ resulting form the original state 
$\rho_{ABE}$ after the local Alice measurement associated with the basis ${\cal B}_{A}$.
In full analogy we have 
\begin{equation}
K_{DW}^{{\cal B}_{AB}}(\rho_{ABE})=I_{A:B}(\rho_{AB}^{CC}) - I_{A:E}(\rho_{AE}^{CQ})
\label{K-CCQ}
\end{equation}
with the suitable reductions of the state $\rho_{ABE}^{CCQ,{\cal B}_{AB}}$ resulting 
form the original state $\rho_{ABE}$ after the product of the two local Alice and Bob 
measurements corresponding to the bases ${\cal B}_{AB}$.
The role of the function $f$ is played just by the mutual information function $I$.

\subsection{General symmetry rule and its simple application}
\label{seccqq}

In what follows we shall use the notation $\hat{U}(X)=UXU^{\dagger}$ 
and $\hat{M}(\{ P_k \})(X)=\sum_{k} P_k X P_k$ with a projectors $P_{k}=P(e_k)=|e_{k}\rangle \langle e_{k} |$
for any orthonormal basis $\{ e_{k} \}$.
We have a simple 

{\it Observation .-} Consider a function $f$ defined on any CQ state on the composite system $XY$
\begin{equation}
\sigma_{XY}^{CQ}=\sum_k P_k \otimes \sigma_k
\end{equation}
Assume that the function $f$ is invariant under some subgroup ${\cal R}_{X}$ of unitary 
operations $R_{X}\in {\cal R}_{X}$ on the system $X$ ie. 
\begin{equation}
\forall_{R_{X}\in {\cal R}_{X}} f(\hat{R}_{X} \otimes \hat{I}_{Y}(\sigma_{XY}^{CQ})=f(\sigma_{XY}^{CQ})
\end{equation}
Given any state $\rho_{XY}$ which is invariant in an analogous way
\begin{equation}
\forall_{R_{X}\in {\cal R}_{X}} \hat{R}_{X} \otimes \hat{I}_{Y}(\rho_{XY})= \rho_{XY}
\end{equation}
we have the following identity
\begin{equation}
f( [\hat{M}_{X}(\{ \hat{R}_{X}(P_k) \}) \otimes \hat{I}_{Y}](\rho_{XY}))=f( [\hat{M}_{X}(\{ P_k \}) \otimes \hat{I}_{Y}](\rho_{XY}))
\end{equation}
for all $R_{X} \in {\cal R}_{X}$ and all $\{ P_{k} \}$ constructed from any orthonormal bases 
$\{ e_k \}$.

{\it Proof. -} It is obvious to see that ,,internal'' $\hat{R}_{X}^{\dagger}$ is absorbed by the 
$QQ$ state $\rho_{XY}$ while the external conjugated one $\hat{R}_{X}$ is absorbed by the invariance of the function $f$.

We have immediate conclusion:

{\it Conclusion.-} The functions (\ref{K-CQQ}) and (\ref{K-CCQ}), if calculated on 
a given generalised private state (\ref{p-bit-mixed}), are invariant under 
the rotations $\hat{R}_{A} \otimes \hat{I}_{BE}$ and $\hat{R}_{A} \otimes \hat{R}_{B} \otimes \hat{I}_{E}$
respectively where $R_{A}$, $R_{B}$ are any unitary operations which are diagonal 
in the local bases $|i\rangle_{A}$, $|j\rangle_{B}$ forming a secure basis of the 
states (\ref{p-bit-mixed}).

{\it Proof .-} The role of the pair of the subsystems 
$\{X,Y\}$ is played by $\{A,BE\}$ or $\{AB,E\}$ respectively 
and the role of the subgroup are all the unitary operations diagonal in the 
bases described in the conclusion.

\subsection{Consequences}
\label{sub:co}

In the case, when the key part dimension $d=2$, this feature can be interpreted graphically. Let us consider CQQ case. Then Alice can choose two angels $(\theta, \varphi)$ to determine her measurement basis using the eigenvectors of the $\sigma_{\hat{n}}$ operator
\begin{gather}
 \left| e_{0} \left( \theta, \varphi\right) \right\rangle = \begin{bmatrix} 
   \cos \frac{\theta}{2} \\
  e^{\mathrm{i} \varphi} \sin \frac{\theta}{2}
\end{bmatrix}, 
 \left|e_{1} \left( \theta, \varphi \right)  \right\rangle = \begin{bmatrix} 
  \sin \frac{\theta}{2}\\  
  -e^{\mathrm{i} \varphi}\cos \frac{\theta}{2}
\end{bmatrix}.
\label{eq:sbv}
\end{gather}
We define the function $K_{D}(\theta$,$\varphi)=K^{DW}_{D} \left( \rho^{C\left(\theta, \varphi\right)QQ}_{priv} \right)$. Here superscript $C(\theta, \varphi)$ denotes that to calculate CQQ state base vectors determined by angels $(\theta, \varphi)$ were used. In spherical coordinate system, in which $\left| e_{0} \left( 0, 0\right) \right\rangle, \left| e_{1} \left( 0, 0\right) \right\rangle$ coincide with the base vectors used in (\ref{p-bit-mixed}) the function $K_{D}(\theta$,$\varphi)$ is $\varphi$ independent, i.e. becomes function only of $\theta$ angle. 
\section{Procedure}
\begin{figure}
	\centering
		\includegraphics[width=80mm]{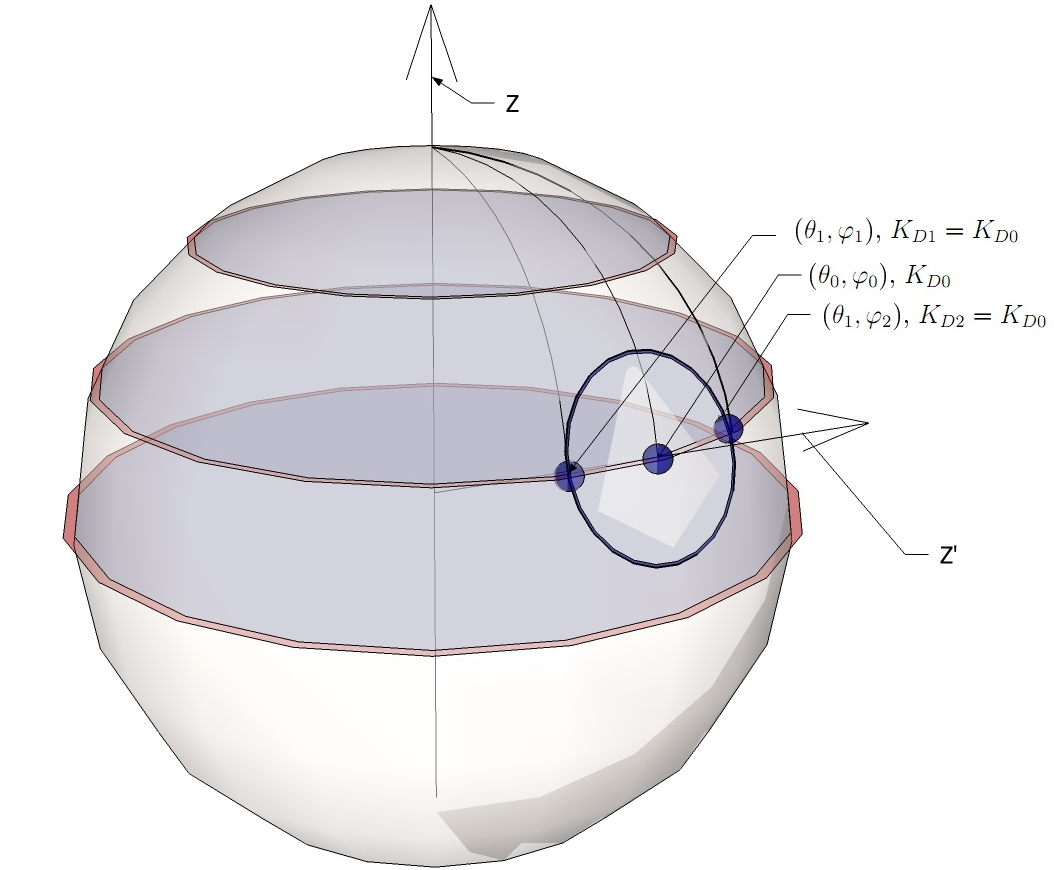}
	\caption{Local Alice's sphere with two different coordinate systems. Z axis corresponds to $\theta_{Max}, \; \varphi_{Max}$ angles whereas Z' to $\theta_{0}, \; \varphi_{0}$ angles. Black circle shows the path along which angle $\varphi$ changes. At the intersection point between Z axis and the sphere $K_{D}(\theta_{Max},\varphi_{Max}) = K_{DMax}$. The intersection point between Z' axis and the sphere is denoted by $(\theta_{0},\varphi_{0})$. At this point $K_{D}(\theta_{0},\varphi_{0}) = K_{D0}$. Points with the same value of $K_{D}$ (laying on a circle, whose center is located at the point $K_{D0}$) are denoted by $(\theta_{1},\varphi_{1})$ and $(\theta_{1},\varphi_{2})$. At the first point $K_{D}(\theta_{1},\varphi_{1}) = K_{D1}$, whereas at the second point $K_{D}(\theta_{1},\varphi_{2}) = K_{D2}$.}
	\label{obrazproc}
	\end{figure}

As follows from the previous section, points on the sphere\footnote{Points on the sphere correspond to angles used in (\ref{eq:sbv}). The term sphere should not be confused with the Bloch sphere of the state.} possessing the same vale of $K_{D}$ establish a circle, whose center is located at intersection of Z axis and the sphere. Moreover, each circle has a center in the same point (all circles are concentric). At this point vale of $K_{D}$ is maximal. 
Using facts presented in Subsection \ref{sub:co} one is able to find such angles $\theta_{Max}$, $\varphi_{Max}$, for which the measurement in basis 
given by (\ref{eq:sbv})
will lead to the maximal value of $K_{D}$, without a priori knowledge of this basis or rotations by which the state was changed. Suppose that the original ideal state was rotated by unknown transformation $U_{A} \otimes I_{BA'B'}$ which eventually changed its optimal measurement basis on Alice side from $\left\{\left|0_{\hat{z}}\right\rangle, \left|1_{\hat{z}}\right\rangle\right\}$ to $\left\{\left|0_{\hat{n}}\right\rangle, \left|1_{\hat{n}}\right\rangle\right\}$ where we define  $\hat{n} = O\hat{z}$ as:
\begin{equation}
\begin{split}
&\left|0_{\hat{n}}\right\rangle\left\langle 0_{\hat{n}} \right| = U\left|0_{\hat{z}}\right\rangle\left\langle 0_{\hat{z}} \right|U^{\dagger} = \frac{1}{2}( I +(O\hat{z})\vec{\sigma}) \\
&\left|1_{\hat{n}}\right\rangle\left\langle 1_{\hat{n}} \right| = U\left|1_{\hat{z}}\right\rangle\left\langle 1_{\hat{z}} \right|U^{\dagger} = \frac{1}{2}(I -(O\hat{z})\vec{\sigma}).
\end{split}
\end{equation}
The procedure is as follows. 
\begin{enumerate}
	\item First one chooses arbitrary values of angles $\theta_{0}$, $\varphi_{0}$ and establishes two base vectors $ \left| e_{0} \left( \theta_{0}, \varphi_{0} \right)  \right\rangle  $ and $ \left| e_{1} \left( \theta_{0}, \varphi_{0} \right) \right\rangle$. This basis is used to perform measurement and to obtain value of $K_{D}$($\theta_{0}$,$\varphi_{0}$) equal $K_{D0}$.
	\item Then one changes the value of $\theta_{0}$ to $\theta_{1}$ and creates a set of base vectors $\left\{ \left| e_{0} \left( \theta_{1} , \varphi_{i} \right)  \right\rangle, \left| e_{1} \left( \theta_{1} , \varphi_{i} \right) \right\rangle \right\}^{N}_{i=1}$. Vectors from this set differ in the value of $\varphi_{i}$ angle by arbitrary constant factor $\frac{2\pi}{N}$ so that $0\leq \varphi_{i} < 2\pi $. One can ascribe each vector from the set to a corresponding point on the sphere. These points lay on a circle, whose centre is located at the point ascribed to vector $ \left| e_{0} \left( \theta_{0}, \varphi_{0} \right) \right\rangle$.
	\item   Subsequently, using the vectors from the set,  the measurements are performed and for each pair of vectors $\left| e_{0} \left( \theta_{1} , \varphi_{i} \right)  \right\rangle, \left| e_{1} \left( \theta_{1} , \varphi_{i} \right) \right\rangle$ the  value of $K_{Di}$ is calculated. A set of values \textit{$K_{D,N}$} =$\left\{K_{D1},\: \ldots,\: K_{DN}\right\}$ is created. The aim of these measurements is to find two points laying on a chosen circle, characterized by values of $\varphi_{i}$ angle, for which value of $K_{Di}$ is equal to earlier calculated value $K_{D0}$. It is not difficult to see that it is always possible to achieve this purpose when we assume continuity of $\varphi$ (or arbitrary small resolution in $\varphi_{i}$). According Subsection \ref{sub:co}, because the sphere is covered with circles with the same value of $K_{D}$, any other circle laying on the sphere can have 0, 1, 2 or infinity intersection points. Thus it is always possible to find such values of $\theta_{0}$, $\theta_{1}$, which ensure that points with the same value of $K_{D}$ are found. Just as for a plane, three points on the sphere are enough to unambiguously determine the circle. The radius and the centre of the circle are found solving the system of equations:
\begin{eqnarray}
\label{soe:ic}
&& d(\theta_{Max} , \varphi_{Max},\theta_{0} , \varphi_{0} ) = R \\
&& d(\theta_{Max} , \varphi_{Max},\theta_{1} , \varphi_{1} ) = R \\
&& d(\theta_{Max} , \varphi_{Max},\theta_{1} , \varphi_{2} ) = R,
\end{eqnarray}  
where \textit{d} is spherical distance defined as \cite{22}:
\begin{equation}
\label{eq:sdd}
d = \arccos (\textbf{P}\cdot\textbf{Q}),
\end{equation} 
here \textbf{P}, \textbf{Q} are two points on the sphere characterized by angles $\theta_{i} , \varphi_{i}$ and $\theta_{j} , \varphi_{j}$, respectively. In therms of Cartesian coordinates ($x = \sin \theta \cos \varphi$,  $y = \sin \theta \sin \varphi $ and $z = \cos \theta$) expression (\ref{eq:sdd}) is of a form:
\begin{eqnarray}
&&d(\theta_{i} , \varphi_{i},\theta_{j} , \varphi_{j} ) = \nonumber  \\ &&= \arccos  (\sin\theta_{i} \sin\theta_{j} \cos(\varphi_{i}-\varphi_{j}) +  \cos\theta_{i} \cos\theta_{j}).
\end{eqnarray}
\end{enumerate}
According to Subsection \ref{sub:co} centre of the circle determined in this way is associated with the basis (characterized by angles $\theta_{Max}$, $\varphi_{Max}$), in which $K_{D} \left( \theta, \varphi \right)$ has maximal value. Fig. \ref{obrazproc} presents main ideas of the proposed procedure. \\ \\ The proposed procedure can be slightly modified. Finding two points with value of $K_{Di}$ exactly equal $K_{D0}$ can cause a problem and such solution is not a practical one. To overcome this difficulty, instead of finding two points with the same value of $K_{D}$, one finds points $K_{D1}$, $K_{D2}$ from the set $K_{D,N}$, for which the values of $K_{D1}, K_{D2}$ are the closest to the $K_{D0}$ i.e. for which $\Delta K_{D1} = K_{D0} - K_{D1} $ and $\Delta K_{D2} = K_{D0} - K_{D2}$ are minimal. Subsequently the interpolating function $I_{\left\{K_{D,N}\right\}}\left(\theta_{1}, \varphi\right)$ from the set $K_{D,N}$ is created. To construct the interpolation function Hermite polynomials of a required order are used. Thus one can write for $i=\left\{1,2\right\}$: 
\begin{equation} \label{taylor}
\begin{split}
K_{D0} &= \\ &= K_{Di} + \Delta K_{Di} =\\ &= K_{D} \left( \theta_{1} , \varphi_{i} \right)+ 
 \left. \frac{\partial I_{\left\{K_{D,N}\right\}}\left(\theta_{1}, \varphi\right)}{\partial \varphi } \right|_{\varphi = \varphi_{i}}  \Delta \varphi_{i} \\ &\quad +  \left. \frac{\partial^{2} I_{\left\{K_{D,N}\right\}}\left(\theta_{1}, \varphi\right)}{\partial \varphi^{2} } \right|_{ \varphi = \varphi_{i}}  \Delta \varphi^{2}_{i} = \\ &= I_{\left\{K_{D,N}\right\}}\left(\theta_{1}, \varphi_{i}\right) + \left. \frac{\partial I_{\left\{K_{D,N}\right\}}\left(\theta_{1}, \varphi\right)}{\partial \varphi } \right|_{\varphi = \varphi_{i}}  \Delta \varphi_{i} \\ &\quad +  \left. \frac{\partial^{2} I_{\left\{K_{D,N}\right\}}\left(\theta_{1}, \varphi\right)}{\partial \varphi^{2} } \right|_{ \varphi = \varphi_{i}}  \Delta \varphi^{2}_{i},
\end{split}
\end{equation}  
where we use the fact that $I_{\left\{K_{D,N}\right\}}\left(\theta_{1}, \varphi_{i}\right) = K_{Di}$ (i.e. the interpolation function reproduces the values of $K_{Di}$ from the set $K_{D,N}$ in the probe points). One solves equation (\ref{taylor}) for $\Delta \varphi_{i}, \; i=\left\{1,2\right\}$.
In general, equation (\ref{taylor}) can have two different solutions. However, in such a case one chooses smaller $\Delta \varphi_{1}$ and $\Delta \varphi_{2}$ (because equation (\ref{taylor}) is Taylor expansion of function $I_{\left\{K_{D,N}\right\}}\left(\theta_{1}, \varphi\right)$ near $\varphi = \varphi_{i}$). By solving modified systems of equations:
\begin{eqnarray}
&& d(\theta_{Max} , \varphi_{Max},\theta_{0} , \varphi_{0} ) = R \\
&& d(\theta_{Max} , \varphi_{Max},\theta_{1} , \varphi_{1} + \Delta \varphi_{1}) = R \\
&& d(\theta_{Max} , \varphi_{Max},\theta_{1} , \varphi_{2} + \Delta \varphi_{2}) = R ,
\label{eq:ms}
\end{eqnarray}  
one obtains values of $\theta_{Max}, \varphi_{Max}$.\\The proposed approach enables to find the basis optimizing the value of $K_{D}$ by performing only local measurements. 
\section{Error estimation}
\begin{figure}
	\centering
		\includegraphics[width=80mm]{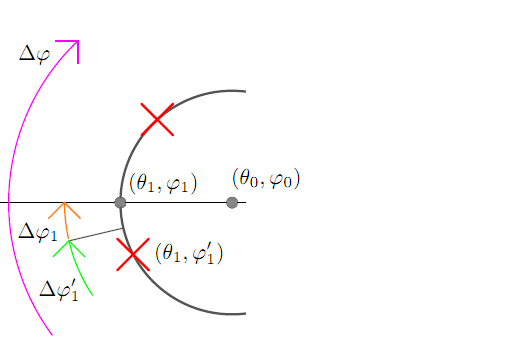}
	\caption{Starting point for the procedure $\left(\theta_0, \varphi_0\right)$, in which $K \left(\theta_0, \varphi_0\right) = K_{D0}$. Point $\left(\theta_1, \varphi_1'\right)$, $K_{D0}=K\left(\theta_1, \varphi_1'\right)$ is closest to the point $ \left( \theta_0, \varphi_0 \right)$ in the given set because$ \Delta K_{D1}=\left|K_{D0}-K_{D1}\right|	$ is minimal  (see text for details). Using the interpolation function one obtains that $K\left(\theta_1, \varphi_1' + \Delta \varphi_1' \right) = K_{D0}$, however, due to numerical error, in fact $\varphi_1' + \Delta \varphi_1' \neq \varphi_1$, where the equality $K_D\left(\theta_1, \varphi_1' + \Delta \varphi_1' \right)=K \left(\theta_0, \varphi_0\right)$ really holds. The case of the second point $\left(\theta_1, \varphi_2' + \Delta \varphi_2' \right)$ is similar. The error of the procedure is given by (\ref{eq:oe}).     }
	\label{erest}
	\end{figure}

Due to approximation (finite sum) and possible numerical errors, it is never possible to solve (\ref{taylor}) exactly. As a result angles $\theta_M', \varphi_M'$ will not lead to the maximal value of distillable key. In this section the estimation of this error is provided.   
Let us denote (see Fig. \ref{erest}):																																																			
\begin{eqnarray}
&& \tilde{\varphi}_1=\varphi_1 + \Delta \varphi_1 = \varphi_1' + \Delta \varphi_1' \nonumber \\ 
&& \tilde{\varphi}_2=\varphi_2 + \Delta \varphi_2 = \varphi_2' + \Delta \varphi_2' \nonumber \\
&& \theta_M + \Delta \theta_M = \theta_M' \nonumber \\
&& \varphi_M + \Delta \varphi_M = \varphi_M'.
\end{eqnarray} Without loss of generality, we can arrange $\tilde{\varphi}_1, \tilde{\varphi}_2$ so that $\tilde{\varphi}_1 > \tilde{\varphi}_2$.  
Spherical distance between points characterized by angles $(\theta_i, \varphi_i)$, $(\theta_j,\varphi_j)$  is given by (\ref{eq:sdd}).
We assume that $(\theta_i, \varphi_i)=(0,0)$, so
\begin{eqnarray}
&& \arccos \left[\cos \theta_M'\right] = R
\label{soe:sd1}
 \\
&& \arccos  \left[\sin \theta_1 \sin \theta_M' \cos \left( \tilde{\varphi}_1-\varphi_M' \right) \right. \left. + \cos \theta_1 \cos \theta_M' \right]  = R
\label{soe:sd2}
 \\
&& \arccos \left[\sin \theta_1 \sin \theta_M' \cos \left( \tilde{\varphi}_2-\varphi_M' \right) \right. + \left. \cos \theta_1 \cos \theta_M' \right] = R.
\label{soe:sd3}
\end{eqnarray} 
Combining \ref{soe:sd2} and \ref{soe:sd3} we get:
\begin{equation}
\cos \left(\tilde{\varphi}_1-\varphi_M'\right) = \cos\left(\tilde{\varphi}_2-\varphi_M'\right).
\end{equation}
Because $\tilde{\varphi}_1 \neq \tilde{\varphi}_2$ and $\varphi_M' \in \left(0, 2 \pi\right] $  there are two possibilities:
\begin{math}
\tilde{\varphi}_1-\varphi_M' = -\left(\tilde{\varphi}_2-\varphi_M'\right)
\end{math} or
\begin{math}
\tilde{\varphi}_1-\varphi_M' = -\left(\tilde{\varphi}_2-\varphi_M' - 2 \pi \right).
\end{math}
We set $\varphi_M':= \varphi_M' \  mod \  2 \pi $
\begin{equation}
\begin{split}
& \varphi_M' = \frac{\tilde{\varphi}_1 + \tilde{\varphi}_2}{2} \\
& \varphi_M + \Delta \varphi_M = \frac{\varphi_1 + \varphi_2}{2} + \frac{ \Delta \varphi_1 + \Delta \varphi_2}{2},
\end{split}
\end{equation} 
so 
\begin{math}
\Delta \varphi_M =  \frac{ \Delta \varphi_1 + \Delta \varphi_2}{2}.
\end{math}
However, we know only $\varphi_i', \ \Delta \varphi_i'$ but we can estimate (see Fig. \ref{erest}) as
\begin{math}
\left|\Delta \varphi_i\right| < \Delta \varphi - \left|\Delta \varphi_i' \right|.
\end{math}  As a result
\begin{equation}
\Delta \varphi_M < \frac{ 2 \Delta \varphi - \left|\Delta \varphi_1' \right| - \left|\Delta \varphi_2' \right|}{2}.
\label{eq:pher}
\end{equation}
Combining equations (\ref{soe:sd1}) and (\ref{soe:sd2})
\begin{equation}
\cot \theta_M' = \cot \frac{\theta_1}{2} \cos \left(\tilde{\varphi}_1 - \varphi_M' \right).
\end{equation}  
As a consequence of the equality
\begin{eqnarray}
&& \cos \left( \tilde{\varphi}_1 - \varphi_M'  \right)  = \cos  \left( \varphi_1 + \Delta \varphi_1 - \frac{\varphi_1 + \Delta \varphi_1 + \varphi_2 + \Delta \varphi_2}{2} \right) = \nonumber \\ &&  \cos  \left(  \frac{\varphi_1 + \Delta \varphi_1 - \varphi_2 - \Delta \varphi_2}{2} \right), 
\end{eqnarray} we obtain the following relation
\begin{equation}
\theta_M = \arccot \left[ \cot \frac{\theta_1}{2} \cos \left(\frac{\varphi_1 - \varphi_2}{2}\right)\right]
\label{eq:th}
\end{equation}
and
\begin{equation}
\theta_M + \Delta \theta_M' = \arccot \left[ \cot \frac{\theta_1}{2} \cos \left(\tilde{\varphi}_1 - \varphi_M' \right)  \right].
\label{eq:ther}
\end{equation}
In order to obtain the upper bound on $\Delta \theta_M'$ we have to find $ \bar{\theta}_M$ - an lower bound on $\theta_M$. Then the following relation holds:
\begin{equation}
\bar{\theta}_M + \Delta \theta_M' < \theta_M + \Delta \theta_M' = \arccot \left[ \cot \frac{\theta_1}{2} \cos \left(\tilde{\varphi}_1 - \varphi_M' \right)  \right],
\label{eq:there}
\end{equation} 
so 
\begin{math}
\Delta \theta_M' < \arccot \left[ \cot \frac{\theta_1}{2} \cos \left(\tilde{\varphi}_1 - \varphi_M' \right)  \right] - \bar{\theta}_M
\end{math}. 
We have to estimate the difference $\varphi_1 - \varphi_2$ using known quantities $\tilde{\varphi}_1, \tilde{\varphi}_2$.  There are two different possibilities: in the first one $\tilde{\varphi}_1 - \tilde{\varphi}_2 < \pi$ whereas in the second $\tilde{\varphi}_1 - \tilde{\varphi}_2 > \pi$. Let us consider the first one. Because $\arccot x \in (-\frac{\pi}{2}, \frac{\pi}{2}]$ is a decreasing function for $x \in (-\infty,0) \cup (0,\infty)$, in order to find $ \bar{\theta}_M$ we have to increase  $\cos \left(\frac{\varphi_1 - \varphi_2}{2}\right)$. For our purposes we shall assume the worst case, namely $\tilde{\varphi}_1 < \varphi_1$ and $\tilde{\varphi}_2 > \varphi_2$. Then $\varphi_1 - \varphi_2 > \tilde{\varphi}_1 - \tilde{\varphi}_2 $. From previous considerations the following relation holds: $\tilde{\varphi}_i + \Delta \varphi > \varphi_i > \tilde{\varphi}_i - \Delta \varphi_i$. As a result $ \tilde{\varphi}_1 - \tilde{\varphi}_2 + 2 \Delta \varphi > \varphi_1 - \varphi_2 > \tilde{\varphi}_1 - \tilde{\varphi}_2$. Using this inequality we get
\begin{equation}
\theta_M >  \bar{\theta}_{M1} =   \arccot \left[ \cot \frac{\theta_1}{2} \cos \left( \frac{\tilde{\varphi}_1 - \tilde{\varphi}_2 + 2 \Delta \varphi}{2}\right) \right].
\label{eq:therpi1}
\end{equation}      
If $\tilde{\varphi}_1 - \tilde{\varphi}_2 > \pi$ the similar line of reasoning leads to  
\begin{equation}
\theta_{M} >  \bar{\theta}_{M2} =   \arccot \left[ \cot \frac{\theta_1}{2} \cos \left( \frac{\tilde{\varphi}_1 - \tilde{\varphi}_2 - 2 \Delta \varphi}{2}\right) \right].
\label{eq:therpi2}
\end{equation}  
Finally one gets
\begin{equation}
\Delta \theta_M' < \arccot \left[ \cot \frac{\theta_1}{2} \cos \left(\tilde{\varphi}_1 - \varphi_M' \right)  \right] - \bar{\theta}_{Mi},
\label{eg:ther}
\end{equation}
where $\bar{\theta}_{Mi}, i=\left\{1,2\right\}$ is given by (\ref{eq:therpi1}) or (\ref{eq:therpi2}). Using perturbed points $(\theta_i,\tilde{\varphi}_i)$ one obtains the point $(\theta_M + \Delta \theta_M,\varphi_M + \Delta \varphi_M)$which differs from the real point $(\theta_M,\varphi_M)$ by $(\Delta \theta_M, \Delta \varphi_M )$, where $\Delta \theta_M, \Delta \varphi_M $ are given by (\ref{eq:ther}) and (\ref{eq:pher}). In the new coordinate system associated with the point $(\theta_M + \Delta \theta_M,\varphi_M + \Delta \varphi_M)$ the error is given by:
\begin{eqnarray}
\label{eq:oe}
\Delta \theta =   \arccos \left[ \right. &&\left. \sin \theta_M' \sin \left( \theta_M' - \Delta \theta_M \right) \cos \Delta \phi_M \right. + \nonumber \\ && + \left. \cos \theta_M' \cos \left( \theta_M' - \Delta \theta_M \right)  \right].
\end{eqnarray}   
\section{Conditions for invaraince of key rate in case
of local action of Pauli channels}
\label{se:ic}
According to \cite{hhho2009}, using appropriate unitary operation $U=1_A \otimes \sum_{i} \left|i\right\rangle\left\langle i  \right| \otimes U^{(i)}_{A'B'}$  (called twisting) it is possible to write a particular private state as:
\begin{equation}
\rho_{priv}=\left|\Psi_+\right\rangle\left\langle \Psi_+ \right|_{AB} \otimes \sigma_{A'B'},
\label{eg:pts}
\end{equation}
where $\left|\Psi_+\right\rangle$ is one of the four Bell states $\left|\Psi_\pm \right\rangle, \ \left|\phi_\pm \right\rangle$
\begin{equation}
\label{bs}
\left|\Psi_\pm \right\rangle = \frac{\left|00\right\rangle \pm \left|11\right\rangle}{\sqrt{2}},\ \ \left|\phi_\pm \right\rangle = \frac{\left|01\right\rangle \pm \left|10\right\rangle}{\sqrt{2}}.
\end{equation}
After sending (\ref{eg:pts}) down the channel $\Lambda_A \otimes I$ where $\Lambda_A(\cdot) = \sum_i p_i K_i (\cdot) K^{\dagger}_i$ with $K_i=\left\{I,\sigma_x,\sigma_y,\sigma_z\right\}$ (note that twisting commutes with the action of the channel) one obtains a state
\begin{equation}
\begin{split}
\tilde{\rho}_{priv}= & \left( p_1 \left|\Psi_+\right\rangle\left\langle \Psi_+ \right|_{AB}  + p_4\left|\Psi_-\right\rangle\left\langle \Psi_- \right|_{AB}  \right. \\ 
 & \left. p_2 \left|\phi_+\right\rangle\left\langle \phi_+ \right|_{AB} + p_3\left|\phi_-\right\rangle\left\langle \phi_- \right|_{AB} \right) \otimes \sigma_{A'B'}.
\end{split}
\end{equation}
The purification of this state is given by
\begin{equation}
\begin{split}
\left|\tilde{\Psi}\right\rangle_{priv}= & \left[\frac{p_1}{\sqrt{2}} \right.  \left(\left|00\right\rangle_{AB} + \left|11\right\rangle_{AB} \right)\left|0\right\rangle_{\bar{E}} \\
&\frac{p_4}{\sqrt{2}}\left(\left|00\right\rangle_{AB} - \left|11\right\rangle_{AB}\right)\left|1\right\rangle_{\bar{E}} \\
&\frac{p_2}{\sqrt{2}}\left(\left|01\right\rangle_{AB} + \left|10\right\rangle_{AB} \right)\left|2\right\rangle_{\bar{E}} \\
& \left. \frac{p_3}{\sqrt{2}}\left(\left|01\right\rangle_{AB} - \left|10\right\rangle_{AB} \right)\left|3\right\rangle_{\bar{E}} \right] \otimes \left| \psi_{\sigma} \right\rangle_{A'B'E} ,
\end{split}
\end{equation}
where $\bar{E}, E$ denote Eves' subsystem. It follows from (\ref{K-CQQ}) that we can trace over subsystems A'B'E (due to additivity of Von Neumann entropy for tensor product states $K_{DW}^{{\cal B}_{A}}$ is independent of subsystems A'B'E).  In order to find the value of the key rate due to Devetak - Winter protocol (\ref{K-CQQ}) we calculate the cqq state using base vectors defined by (\ref{eq:sbv}). The nonzero elements of the reduced AB matrix are given by
\begin{equation}
\begin{split}
&\frac{1}{2} a_{0000}= \cos^2 \frac{\theta}{2} \left(p_1 + p_4 \right) + \sin^2 \frac{\theta}{2} \left(p_2 + p_3 \right) \\
&\frac{1}{2} a_{0101}= \sin^2 \frac{\theta}{2} \left(p_1 + p_4 \right) + \cos^2 \frac{\theta}{2} \left(p_2 + p_3 \right) \\
&\frac{1}{2} a_{1010}= \sin^2 \frac{\theta}{2} \left(p_1 + p_4 \right) + \cos^2 \frac{\theta}{2} \left(p_2 + p_3 \right) \\
&\frac{1}{2} a_{1111}= \cos^2 \frac{\theta}{2} \left(p_1 + p_4 \right) + \sin^2 \frac{\theta}{2} \left(p_2 + p_3 \right) \\
&\frac{1}{2} a_{0001}=a^*_{0100}=e^{i \varphi}\sin \frac{\theta}{2} \cos \frac{\theta}{2} \left(p_1 - p_4 \right) + e^{-i \varphi} \sin \frac{\theta}{2} \cos \frac{\theta}{2} \left(p_2 - p_3 \right) \\
&\frac{1}{2} a_{1011}=a^*_{1110}= e^{i \varphi}\sin \frac{\theta}{2} \cos \frac{\theta}{2} \left(p_4 - p_1 \right) +  e^{-i \varphi} \sin \frac{\theta}{2} \cos \frac{\theta}{2} \left(p_3 - p_2 \right).
\end{split}
\end{equation} 
As a result entropies of Alice and Bob are equal $S_A=S_B=1$. The reduced AB matrix is block diagonal so its eigenvalues are
\begin{equation}
\begin{split}
&\lambda_{1,2} = \frac{1}{4} \left(1 + \sqrt{  \cos^2 \theta\left( p_1+p_4  - p_2-p_3 \right)^2 + \sin^2 \theta \left|e^{i \varphi} \left(p_1 - p_4 \right) + e^{-i \varphi}  \left(p_2 - p_3 \right)\right|^2  } \right)   \\
&\lambda_{3,4} = \frac{1}{4} \left(1 - \sqrt{ \cos^2 \theta\left( p_1+p_4  - p_2-p_3 \right)^2 + \sin^2 \theta \left| e^{i \varphi} \left(p_1 - p_4 \right) +  e^{-i \varphi}  \left(p_2 - p_3 \right)  \right|^2 } \right).
\end{split}
\end{equation}
As a consequence, $I_{A:B}$ will be independent of $\varphi$ angle if and only if $p_1-p_4$ or $p_2 =p_3$. In order to minimize joint entropy $S_{A:B}$ one has to set $\theta = 0$ or $\theta = \frac{\pi}{2}$ depending on $\left\{ p_1,p_2,p_4 \right\}$ or $\left\{p_1,p_2,p_3 \right\}$. Setting $\theta = 0$ will be optimal if $(p_1+p_4-2 p_2)>0$ for $p_2=p_3$ or $(p_2+p_3-2 p_1)>0$ for $p_1=p_4$ , otherwise $\theta = \frac{\pi}{2}$ . In order to show that $I_{A:\bar{E}}$ is independent of $\varphi$ and let us consider a state resulting from the measurement performed on $\bar{E}$ subsystem. This operation does not increase the value of $I_{A:\bar{E}}$ (which we denote as $I^M_{A:\bar{E}}$ ) so we have $I_{A:\bar{E}} \geq I^{M}_{A:\bar{E}}$ and due to (\ref{K-CQQ}) 
\begin{equation}
K_{DW}^{{\cal B}_{A}}(\tilde{\rho}_{priv})=I_{A:B} - I_{A:\bar{E}} \geq I_{A:B} - I^{M}_{A:\bar{E}}.
\end{equation}
After measurement the reduced A$\bar{E}$ matrix has following eigenvalues $\lambda_{1,2}=\frac{p_1}{2},\lambda_{3,4}=\frac{p_2}{2},\lambda_{5,6}=\frac{p_3}{2},\lambda_{7,8}=\frac{p_4}{2}$ whereas the eigenvalues of $\bar{E}$ matrix are given by $\lambda_{1}=p_1,\lambda_{2}=p_2,\lambda_{3}=p_3,\lambda_{4}=p_4$
so we obtain that $I_{A:\bar{E}}$ is independent of $\theta$. As a result if $p_1=p_4$ or $p_2=p_3$ the distillable key will preserve its invariance and the proposed procedure will be valid.        
\section{Numerical results}
\begin{figure}[p]
	\centering
		\includegraphics[width=130mm]{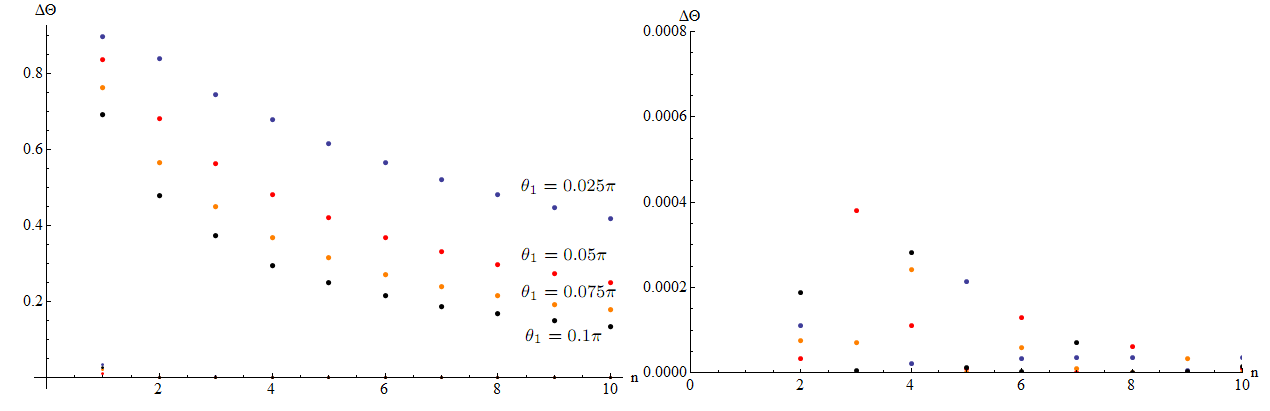}
	\caption{The estimated error of the procedure (using formula (\ref{eq:oe})) - left plot, and the error of the procedure (absolute value of the difference between rotation angle and the angle calculated by the procedure  $\Delta \theta = \left|\theta - \theta_{M'}\right|$ ) -right plot vs. the number of points r=10n (n=1 denotes that 10 points were used) for the rotated $\rho_{SWAP}$ state. Rotation angle $\theta=\frac{\pi}{3}$. Different colors correspond to different choices of $\theta_1$ angle in (\ref{eq:ms}): blue - $\theta_1$ =0.0025 $\pi$, red - $\theta_1$ =0.005 $\pi$, orange - $\theta_1$ =0.0075 $\pi$, red - $\theta_1$ =0.01 $\pi$. In agreement with (\ref{eq:oe}), for all cases the estimated error of the procedure constitute an upper bound on the error. The error of the procedure is orders of magnitude smaller than its estimated value.}
	\label{rotpbit}
	\end{figure}

In this section we provide some examples of the results obtained by implementing the above procedure numerically. 
{\it Rotated $\rho_{SWAP}$ state. - }
Consider a private state introduced in \cite{secret-key-bound} and realized experimentally \cite{dkdbh2011} 
\begin{eqnarray}
\label{pbit}
\rho_{SWAP}=&& \frac{1}{4} \left|\Psi_- \right\rangle \left\langle \Psi_- \right|_{AB} \otimes \left|\Psi_- \right\rangle \left\langle \Psi_- \right|_{A'B'} + \nonumber \\ && 
\frac{1}{4} \left|\Psi_+ \right\rangle \left\langle \Psi_+ \right|_{AB} \otimes  I_{A'B'} \nonumber \\ &&- \frac{1}{4} \left|\Psi_+ \right\rangle \left\langle \Psi_+ \right|_{AB} \otimes  \left|\phi_- \right\rangle \left\langle \phi_- \right|_{A'B'}. 
\end{eqnarray}
This state was rotated and then the optimizing procedure was applied. The example of the results is shown in Fig. \ref{rotpbit}.
 
\begin{figure}[p]
	\centering
		\includegraphics[width=130mm]{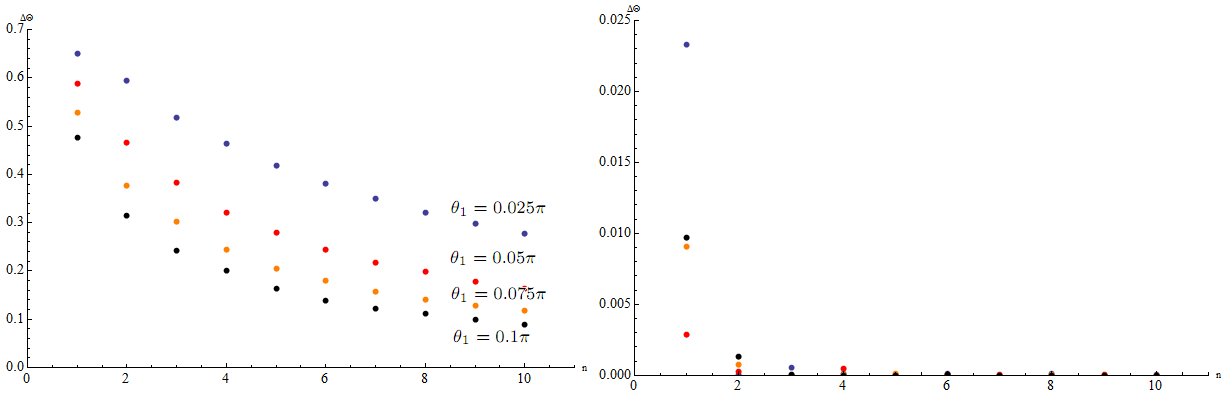}
	\caption{The estimated error of the procedure (using formula (\ref{eq:oe})) - left plot and, the error of the procedure (absolute value of the difference between rotation angle and the angle calculated by the procedure  $\Delta \theta = \left|\theta - \theta_{M'}\right|$ ) -right plot vs. the number of points r=10n (n=1 denotes that 10 points were used) for the state $\tilde{\rho}_{SWAP}=\Lambda_A \otimes I_{BA'B'}$, where $\Lambda(\rho) = p \frac{I}{2} + (1-p)\rho$. Rotation angle $\theta = \frac{\pi}{4}$, $p=\frac{1}{10}$. Different colors correspond to different choices of $\theta_1$ angle in (\ref{eq:ms}): blue - $\theta_1$ =0.0025 $\pi$, red - $\theta_1$ =0.005 $\pi$, orange - $\theta_1$ =0.0075 $\pi$, red - $\theta_1$ =0.01 $\pi$. In agreement with (\ref{eq:oe}), for all cases the estimated error of the procedure constitute an upper bound on the error. The error of the procedure is orders of magnitude smaller than its estimated value.}
	\label{depc}
	\end{figure}

{\it Depolaraizig channel. - }
The procedure was checked using the rotated state $\tilde{\rho}_{SWAP}=\Lambda_A \otimes I_{BA'B'}$, where $\Lambda(\rho) = p \frac{I}{2} + (1-p)\rho$. The example of the results is shown in Fig. \ref{depc}.

\begin{figure}[p]
	\centering
		\includegraphics[width=130mm]{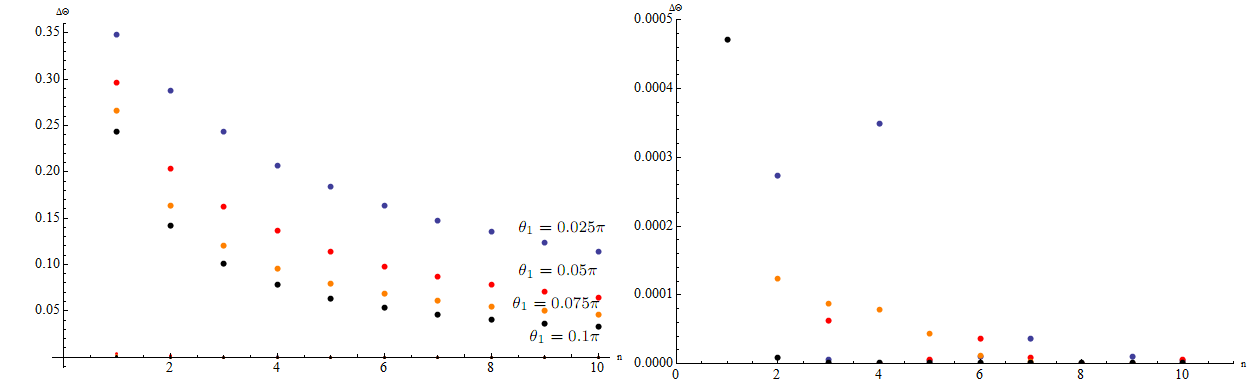}
	\caption{The estimated error of the procedure (using formula (\ref{eq:oe})) - left plot and the error of the procedure (absolute value of the difference between rotation angle and the angle calculated by the procedure  $\Delta \theta = \left|\theta - \theta_{M'}\right|$ ) -right plot vs. the number of points r=10n (n=1 denotes that 10 points were used) for the state $\tilde{\rho}_{SWAP}=\Lambda_A \otimes I_{BA'B'}$, where $\Lambda(\rho) = p \rho + (1-p)\sigma_z \rho \sigma_z$. Rotation angle $\theta = \frac{\pi}{7}$, p=$\frac{3}{10}$. Different colors correspond to different choices of $\theta_1$ angle in (\ref{eq:ms}): blue - $\theta_1$ =0.0025 $\pi$, red - $\theta_1$ =0.005 $\pi$, orange - $\theta_1$ =0.0075 $\pi$, red - $\theta_1$ =0.01 $\pi$. In agreement with (\ref{eq:oe}), for all cases the estimated error of the procedure constitute an upper bound on the error. The error of the procedure is orders of magnitude smaller than its estimated value.     }
	\label{pfc}
	\end{figure}

{\it Phase flip channel. - }
Another test was performed using the rotated $\tilde{\rho}_{SWAP}=\Lambda_A \otimes I_{BA'B'}$, where $\Lambda(\rho) = p \rho + (1-p)\sigma_z \rho \sigma_z$. The example of the results is shown in Fig. \ref{pfc}.

\begin{figure}[p]
	\centering
		\includegraphics[width=130mm]{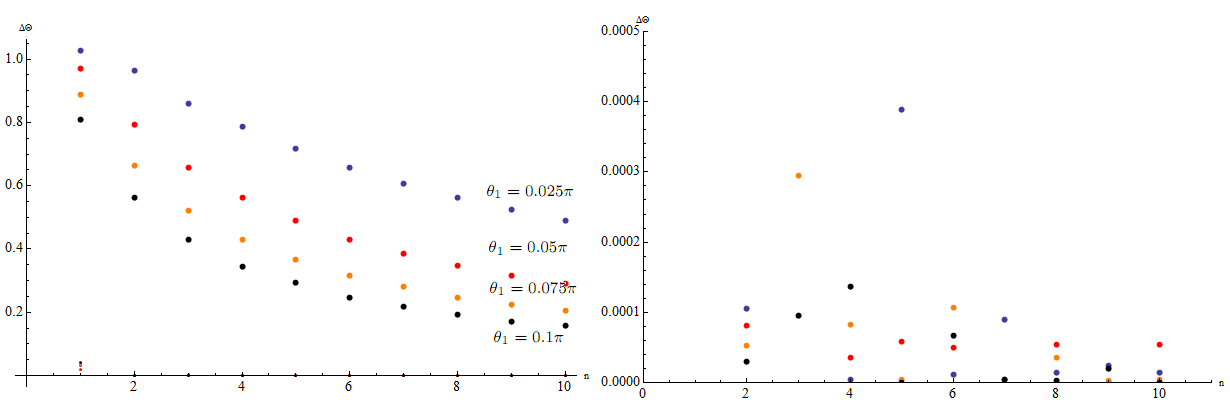}
	\caption{The estimated error of the procedure (using formula (\ref{eq:oe})) - left plot and the error of the procedure (absolute value of the difference between rotation angle and the angle calculated by the procedure  $\Delta \theta = \left|\theta - \theta_{M'}\right|$ ) -right plot vs. the number of points r=10n (n=1 denotes that 10 points were used) for the state $\rho_{priv}=p\rho_{SWAP}+(1-p)\rho_{MSWAP}$. Rotation angle $\theta = \frac{\pi}{8}$, p=$\frac{2}{5}$. Different colors correspond to different choices of $\theta_1$ angle in (\ref{eq:ms}): blue - $\theta_1$ =0.0025 $\pi$, red - $\theta_1$ =0.005 $\pi$, orange - $\theta_1$ =0.0075 $\pi$, red - $\theta_1$ =0.01 $\pi$. In agreement with (\ref{eq:oe}), for all cases the estimated error of the procedure constitute an upper bound on the error. The error of the procedure is orders of magnitude smaller than its estimated value. }
	\label{pbm}
	\end{figure}

{\it Rotated mixture of  $\rho_{SWAP}$ and $\rho_{MSWAP}$  states. - }
Another example of the private states is a state 
\begin{eqnarray}
\label{pbit}
\rho_{MSWAP} =&& \frac{1}{2} \left|\phi_- \right\rangle \left\langle \phi_- \right| \otimes \left( \frac{1}{2} \left|00\right\rangle  \left\langle 00\right| +  \left| \Psi_+ \right\rangle \left\langle  \Psi_+ \right| \right)+ \nonumber \\ && + \frac{1}{2} \left|\phi_+\right\rangle \left\langle \phi_+ \right| \otimes \left(\frac{1}{2} \left|11\right\rangle \left\langle 11\right| \left| \Psi_- \right\rangle \left\langle \Psi_- \right| \right).
\end{eqnarray}
This state plays a role in bound entangled secure key \cite{hphh2008}. We checked the procedure using rotated mixture of two private states $\rho_{priv}=p\rho_{SWAP}+(1-p)\rho_{MSWAP}$ and the rotated $\tilde{\rho}_{SWAP}=\Lambda_A \otimes I_{BA'B'}$, where $\Lambda(\rho) = p \rho + (1-p)\sigma_z \rho \sigma_z$. The example of the results is shown in Fig. \ref{pbm}.
 
\begin{figure}[p]
	\centering
		\includegraphics[width=130mm]{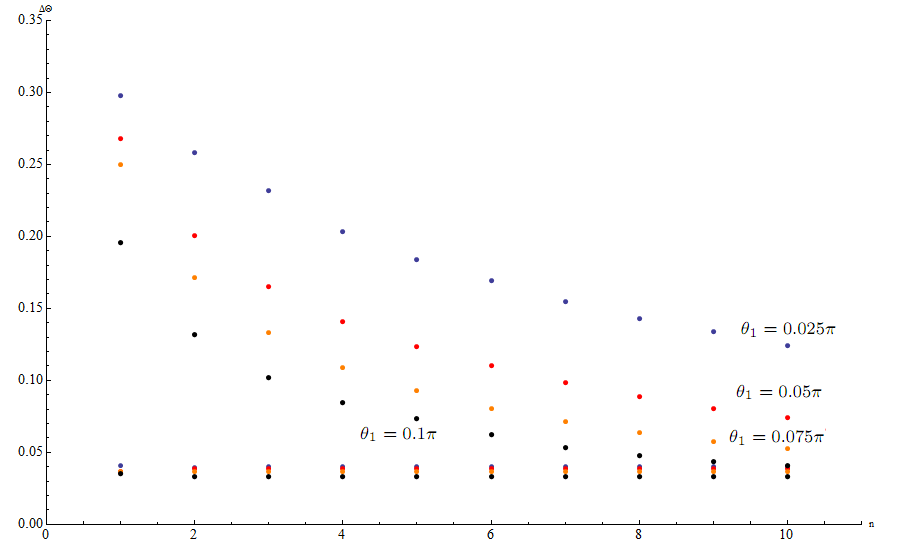}
	\caption{The estimated error of the procedure (using formula (\ref{eq:oe})) and the error of the procedure (absolute value of the difference between rotation angle and the angle calculated by the procedure  $\Delta \theta = \left|\theta - \theta_{M'}\right|$ ) vs. the number of points r=10n (n=1 denotes that 10 points were used) for the state $\rho_{priv}=(\Lambda_A \otimes 1_{BA'B'}) \rho_{SWAP}$, $\Lambda_A(\rho)=K^{\dagger}_1 \rho K_1 + K^{\dagger}_2 \rho K_2$ where $K_i$ are given by (\ref{eq:tp}). Rotation angle $\theta = \frac{\pi}{8}$, u=0.1 $\pi$, v=0.05 $\pi$. Different colors correspond to different choices of $\theta_1$ angle in (\ref{eq:ms}): blue - $\theta_1$ =0.0025 $\pi$, red - $\theta_1$ =0.005 $\pi$, orange - $\theta_1$ =0.0075 $\pi$, red - $\theta_1$ =0.01 $\pi$.  As discussed in Sec.\ref{se:ic} in this case proposed estimation scheme is no longer valid. The error of the procedure can be grater than its estimated value.}
	\label{tc}
	\end{figure}

{\it Qubit channel with trigonometrical parametrization. - }
Consider a channel given by the Kraus operators \cite{rszw2002}
\begin{eqnarray}
\label{eq:tp}
&& K_1 = \left[\cos\frac{u}{2}  \cos \frac{v}{2}\right] I + \left[\sin\frac{u}{2}  \sin \frac{v}{2}\right] \sigma_z \nonumber \\
&& K_2 = \left[\cos\frac{u}{2}  \sin \frac{v}{2}\right] \sigma_x - \mathrm{i} \left[\sin\frac{u}{2}  \cos \frac{v}{2}\right] \sigma_y
\end{eqnarray}
which transforms the Bloch vector $\vec{r} = \left[r_x,r_y,r_z\right]^T$ of the state into $\vec{r'} = \left[\cos u r_x,\cos v r_y,\cos u \cos v r_z + \sin u \sin v \right]^T$. It follows from Section \ref{se:ic} that in general this channel does not preserve the invariance of distillable key. In this case the procedure fails. The example of the results is shown in Fig. \ref{tc}.

\section{Conclusions}
In the present paper the new symmetry of the states with perfect secure key called generalized private states have been provided which says that the most popular Devetak-Winter protocol secret key rate is invariant 
in both scenarios of CCQ and CQQ type if the measurement bases, chosen in a wrong way, 
are rotated (in a sense of angular momentum) around the axis corresponding to the secure basis by any angle. 
The symmetry has a particularly good interpretation when seen on a sphere since then 
the wrong is any basis corresponding to $\hat{n}$ with an angle $\theta$ to the $\hat{z}$ axis 
(corresponding to the secure basis) while the symmetry rotation is just the rotation by the $\phi$ 
angle around the $\hat{z}$ axis. 
We have also proven that for the qubit key part the optimality of the $\hat{z}$ axis as the secure basis is
preserved after the action of any bistochastic channel (i.e. the one represented by random 
Pauli rotations).

The symmetry of the ideal p-bit lead us to the heuristic scheme of estimation of the optimal axis 
(with respect to the Devetak-Winter secret key rate $K_{DW}$) which is valid for any state 
that has this type of symmetry of the key rate under the rotation around the optimal basis.
Namely given the density matrix, may be even in a numerical form, instead 
of searching over all sphere Alice may perform the analysis of the key 
over a ring around some chosen axis $\hat{z}'$ on the sphere and guess the 
optimal measurement axis only on the data based on this ring. 

The method generally has a ,,dualistic'' character with respect 
to the channel action. 
If the Alice subsystem as the direction of the optimal axis unperturbed, 
than the results are good if the $\hat{z}'$ is chosen 
to be far form the (unknown) optimal one $\hat{z}$ while if 
there is a perturbation of the optimal direction $\hat{z}$
(in a sense of the shrinking of that direction on a Bloch sphere)
then the closer is the chosen axis to the original one the result 
is better. 
Basing on the polynomial approximation of the key function 
on the chosen ring there is also the possibility of the 
derivation of the error bar of the procedure. 
The analysis of examples shows that the error bar in general 
bounds the actual value of the error made in the procedure. 

We believe that the present method may be especially useful 
when the large sample of data are provided and quick estimation of optimal 
Alice measurement is needed.

	\noindent {\bf Acknowledgments}. We thank Ewa and J\c{e}drzej Tuziemscy for help in preparation of Figiures. Calculations were carried out at the Academic Computer Center in Gda\'nsk. This work was supported by 7th Framework Programme Future and Emerging Technologies project Q-ESSENCE.

\appendix



\begin{thebibliography}{10}


\bibitem{reviews}
R. Horodecki, P. Horodecki, M. Horodecki and K.
Horodecki, Rev. Mod. Phys. {\bf 81}, 865 (2009) 
\bibitem{BB84}
C. H. Bennett and G. Brassard,  Proceedings of the
IEEE International Conference on Computers, Systems and
Signal Processing IEEE Computer Society, New York, 1984, pp. 175–179.
\bibitem{Ekert}
A. K. Ekert, Phys. Rev. Lett. {\bf 67}, 661 (1991).
\bibitem{BBM}
C. H. Bennett, G. Brassard, and N. D. Mermin, 1992, Phys.
Rev. Lett. {\bf 68}, 557 (1992).
\bibitem{QPA}D. Deutsch, A. Ekert, R. Jozsa, C. Macchiavello, S. Popescu,
and A. Sanpera, Phys. Rev. Lett. {\bf 77}, 2818 (1996).
\bibitem{Bennett-distillation}
C. H. Bennett, G. Brassard, S. Popescu, B. Schumacher, J. A.
Smolin, and W. K. Wootters, Phys. Rev. Lett. {\bf 76}, 722 (1996).
\bibitem{SP2000}
P. W. Shor, J. Preskill, Phys. Rev. Lett. {\bf 85}, 441 (2000) 
\bibitem{bound}M. Horodecki, P. Horodecki, and R. Horodecki, Phys.
Rev. Lett. {\bf 80}, 5239 (1998). 
\bibitem{secret-key-bound}
K. Horodecki, M. Horodecki, P. Horodecki, and J. Oppenheim, Phys. Rev. Lett. {\bf 94}, 160502 (2005).
\bibitem{Dobek-Et-Al}
K. Dobek, M. Karpinski, R. Demkowicz-Dobrzanski, K. Banaszek, P. Horodecki,
Phys. Rev. Lett. {\bf 106}, 030501 (2011). 
     \bibitem{hhho2009} 
  K. Horodecki, M. Horodecki, P. Horodecki, J. Oppenheim 
 IEEE Trans. Inf. Theory {\bf 55}, 1898 (2009).
  \bibitem{dkdbh2011} 
  K. Dobek, M. Karpinski, R. Demkowicz-Dobrzanski, K. Banaszek, P. Horodecki 
  Phys. Rev. Lett. {\bf 106}, 030501 (2011).
 \bibitem{22}
T. Rowland 
   \emph{ Spherical Distance.} 
  MathWorld--A Wolfram Web Resource, created by Eric W. Weisstein. http://mathworld.wolfram.com/SphericalDistance.html  
 \bibitem{hphh2008} 
K. Horodecki, £. Pankowski, M. Horodecki, P. Horodecki 
IEEE Trans. Inf. Theory {\bf 54}, 2621 (2008).
\bibitem{rszw2002}
M. B. Ruskai, S. Szarek, E. Werner
Lin. Alg. Appl. {\bf 347}, 159 (2002).
 \end{thebibliography}
\end{document}